\documentclass[12pt,preprint]{aastex}

\begin{document}

\newcommand\etal{et al. }

\def\etal{et al.\ }
\def\msun{M_{\sun}}
\def\rsun{R_{\sun}}
\def\lsun{L_{\sun}}
\def\kms{\rm \, km \, s^{-1}}
\def\ha{H$\alpha \;$}
\def\ecs{\rm erg \, cm^{-2} \, s^{-1}}
\def\micron{$\mu$m}
\newcommand\mdot{\dot{M}}
\newcommand\msunyr{\rm M_{\odot}\,yr^{-1}}

\title{The Outburst of V1647 Ori Revealed by {\it{Spitzer}}}
\author{J. Muzerolle \altaffilmark{1}, S. T. Megeath \altaffilmark{2},
K. M. Flaherty \altaffilmark{2}, K. D. Gordon \altaffilmark{1},
G. H. Rieke \altaffilmark{1},
E. T. Young \altaffilmark{1}, C. J. Lada \altaffilmark{2}}
\altaffiltext{1}{Steward Observatory, University of Arizona, 933 N. Cherry
Ave., Tucson, AZ 85721 (jamesm@as.arizona.edu)}
\altaffiltext{2}{Harvard-Smithsonian Center for Astrophysics, Mail Stop 42,
60 Garden Street, Cambridge, MA 02138}

\begin{abstract}

We present {\it{Spitzer}} Space Telescope observations of V1647 Ori,
the outbursting source lighting McNeil's nebula, taken near the optical peak
of the outburst in early March 2004.
The source is easily detected in all {\it{Spitzer}} imaging bands
from 3.6 - 70 \micron.  The fluxes at all wavelengths are roughly a factor
of 15 brighter than pre-outburst levels; we measure a bolometric luminosity
of 44 $\lsun$.  We posit that this event is
due to an increase in the accretion luminosity of the source.
Simple models of an accretion disk plus tenuous envelope can qualitatively
explain the observed pre- and post-outburst spectral energy distributions.
The accretion activity implied by our results indicates that the outburst
may be intermediate between FUor and EXor-type events.
We also report the discovery of a previously unknown mid-infrared
counterpart to the nearby Herbig-Haro object HH 22.
\end{abstract}


\section{Introduction}

McNeil's nebula was first reported in late January 2004 as a new optically-
revealed object
in the vicinity of the NGC 2068 star forming region.  Subsequent studies
showed it to be an outburst event that began in November/December 2003,
and rose in brightness by about 4 magnitudes at $I$ over the next few months
(Brice\~no et al. 2004).  The morphology of the nebula
is consistent with a cavity in the envelope of a protostar
scattering light from the highly reddened central source
(Brice\~no et al. 2004; Reipurth \& Aspin 2004; Kun et al. 2004).
The presumed source of the outburst, V1647 Ori, corresponds to
the mid-infrared source IRAS 05436-0007, whose pre-outburst spectral energy
distribution (SED) is that of a flat-spectrum source (Abraham et al. 2004).
Such objects are usually considered to be in transition between the Class I
and II pre-main sequence evolutionary phases, with accretion disks and
dissipating infalling envelopes.  Near-infrared spectral characteristics
showing both disk accretion signatures such as hydrogen and CO emission
(Reipurth \& Aspin 2004; Vacca et al. 2004) and envelope features
such as water and CO ice absorption (Vacca et al. 2004), as well as
the apparent lack of a molecular outflow (Lis et al. 1999), all seem
to confirm the transitionary nature of this object.

The appearance of V1647 Ori has provided a rare opportunity to study
eruptive phenomena in young stellar objects (YSOs), events which are likely
to have a significant effect on the evolution of young stars and
their circumstellar material.
The post-outburst data to date have suggested a significant increase
in luminosity by factors of 10-25 in the optical and near-infrared
(Brice\~no et al. 2004; Reipurth \& Aspin 2004).
Andrews et al. (2004), based on $N$-band photometry
and spectroscopy and submillimeter observations, estimated a total bolometric
luminosity in the range of $\sim 30-90 \; \lsun$.  Spectral features
such as P Cygni profiles at H$\alpha$, Pa$\beta$, and He I indicate
a strong wind associated with the outburst, probably accretion-generated
(Reipurth \& Aspin 2004; Vacca et al. 2004; Walter et al. 2004).
Jet-related features such as forbidden emission lines or
shocked molecular hydrogen are not seen near the source,
although a collimated outflow is suggested by the presence of HH 23
roughly an arcminute to the north.
The nature of the outburst is not yet clear,
as the observational characteristics are not completely consistent with
either of the main classes of eruptive variables seen in YSOs,
the FUors and EXors.  FUors have outburst timescales of decades,
with an initial brightening of typically 4-5 magnitudes in the optical,
while EXor events are generally much shorter (of the order of weeks or months)
and exhibit smaller luminosity increases.

We present serendipitous {\it{Spitzer}} observations of V1647 Ori
and its immediate surroundings.
Our data span a wavelength range of 3.6 - 70 \micron, obtained separately
with the Infrared Array Camera (IRAC) and Multiband Imaging Photometer
for {\it{Spitzer}} (MIPS) within the span of about a week near the peak
of the outburst.
These data provide a unique opportunity to study a rare outburst event
over an unprecedented range of wavelength, probing physical effects
over a large range of dust temperatures and hence
distances from the central source.

\section{Observations}

V1647 Ori was observed by both IRAC and MIPS in the course
of mapping the NGC 2068/2071 star formation region, as part of a GTO
science program.  The IRAC observations were taken on March 7, 2004.
Two dithers were taken at each position, with a 0.4 and 10.4 second
integration time per dither.  The linearization, flat fielding and
dark subtraction of the data were performed by using the IRAC pipeline
at the Spitzer Science Center, version S9.5.  Since the source was
saturated in the first two channels, we determined the fluxes in
those bands by fitting a PSF to the unsaturated part of the data,
and then measuring the magnitude via aperture
photometry of the fitted PSF.  Fluxes for the other two channels
were determined via aperture photometry
of the individual 0.4 second frames, which were not saturated.
The zero points for the absolute flux calibration are
19.660, 18.944, 16.880, and 17.394 in channels 1-4, respectively.
We estimate total uncertainties of about 10\%.

The same region was mapped with MIPS using scan mode (Rieke et al. 2004)
on March 15, 2004.  Medium scan rate with half-array offsets
was employed, providing a total effective exposure
time per pixel of 80 seconds at 24 {\micron} and 40 seconds at
70 {\micron}.  The data were reduced and
mosaicked using the instrument team in-house Data Analysis Tool,
which includes per-exposure calibration of all detector
transient effects, dark current subtraction, and
flat fielding/illumination correction (Gordon et al. 2004).
V1647 Ori is saturated in the PSF core at 24 {\micron}.
We derived photometry by again fitting to the unsaturated part of
the PSF.  We estimate an error of $\sim 10\%$ for this process,
which is comparable to that of the systematic
uncertainty in our overall photometric calibration.

Portions of the final IRAC channel 2 and MIPS 24 {\micron} mosaics
around V1647 Ori are shown in Figure~\ref{24image}.  
Several other objects of interest are seen here.
The bright source to the east of V1647 Ori corresponds to
the T Tauri star LkH$\alpha$ 301.  We also discover
two previously unknown sources roughly 30" to the SW and NE of V1647 Ori,
the latter of which corresponds to Herbig-Haro object
HH 22A.  Most of these sources will be presented in a future
paper on the entire NGC 2068 region; however, we briefly discuss
HH 22A in section 3.

\section{Results}

\subsection{Outburst SED}

\begin{figure}
\epsscale{1.0}
\plotone{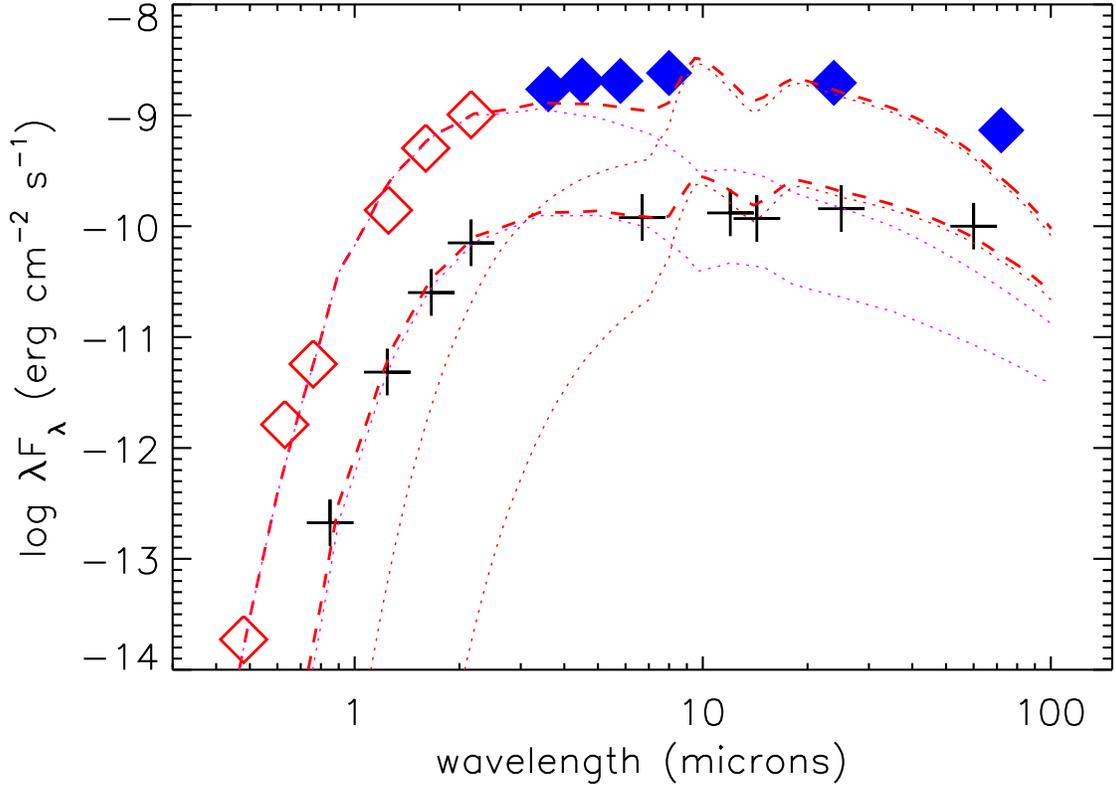}
\caption{Observed and model SEDs for V1647 Ori.  {\it{Plus signs:}}
Pre-outburst SED, data from {\it{2MASS}}, {\it{IRAS}}, Abraham et al. (2004),
Mitchell et al. (2001), and Lis et al. (1999).  {\it{Open diamonds:}}
Outburst photometry from Reipurth \& Aspin (2004).
{\it{Solid diamonds:}} {\it{Spitzer}}
photometry from this work.
{\it{Magenta dotted lines:}} Viscous accretion disk emission, with a disk
inclination of 50$^{\circ}$, mass accretion rates of 6 $\times 10^{-7}$
$\msunyr$ (pre-outburst) and $10^{-5}$ $\msunyr$ (outburst), and visual
extinction 15 (pre-outburst) and 11 (outburst).
{\it{Red dotted lines:}} Infalling envelope emission, with a mass infall
rate of $10^{-6}$ $\msunyr$, inner and outer envelope radii of 1 and 2000
AU, respectively, filling factor of 0.3, and inner temperatures of
350 K (pre-outburst) and 550 K (outburst).
{\it{Dot-dashed lines:}} Sum of the disk and envelope emission.
\label{seds}}
\end{figure}

The resulting post-outburst SED from our {\it{Spitzer}} measurements
is shown in Figure~\ref{seds}.  Also plotted are optical $g'r'i'$ and
near-infrared $JHK'$ measurements from Reipurth \& Aspin (2004).
These data are not simultaneous; the near-infrared data in particular
were taken over a month before the {\it{Spitzer}} data.
Nevertheless, we believe all these measurements should be representative
of the outburst near its maximum, as optical and near-infrared photometry
spanning this range of dates shows little overall change in the source
brightness (e.g., Brice\~no et al. 2004; Kun et al. 2004; McGehee et al. 2004;
Walter et al. 2004).  We do note that the scatter between different
optical studies is quite large, which probably reflects significant
contamination from scattered light, not surprising given the presence
of an envelope around the source (see below).
The SED shape is qualitatively similar to the pre-outburst SED
(also shown in Fig.~\ref{seds}),
with the spectrum increasing sharply to $\sim 2 \mu$m and then
becoming flatter at longer wavelengths.
The overall luminosity increase is impressive,
with a factor of 15-20 increase in brightness
across the spectrum from the optical to 70 $\mu$m.  We measure
a total bolometric luminosity of $44 \; L_{\odot}$, a factor of
$\sim 15$ increase from the pre-outburst state.

\subsection{Models}

The most likely
explanation for this event is an increase in the accretion luminosity
of the central source driven by an instability in the accretion disk,
which subsequently illuminated an envelope 
that surrounds the source.  To explore this idea in more
depth, we will test whether the protostellar characteristics of
an accretion disk plus infalling envelope can explain the observed SED
both before and during the outburst.  We have constructed simple
models of both a viscous accretion disk (Kenyon et al. 1988)
and a remnant optically thin envelope (Adams et al. 1987; Kenyon \&
Hartmann 1991), changing the accretion luminosity produced by
the disk by a factor of $\sim 15$ to simulate the outburst.

We first employ a standard steady viscous accretion disk model,
as first applied to FU Orionis objects
(Kenyon et al. 1988).  In this formulation, the disk temperature is
specified by the mass and radius of the central star, the mass
accretion rate $\mdot$ in the disk, and a radial dependence of $r^{-3/4}$.
The disk is assumed to be flat, with radiation at each radius given
by a blackbody at the appropriate temperature.  We adopted a typical
T Tauri stellar mass and radius of $M=0.5 \; M_{\odot}$ and
$R=2 \; R_{\odot}$, as suggested by the low pre-outburst bolometric luminosity.
For the outburst spectrum, we derived a disk mass accretion rate
of $\sim 10^{-5} \; \msunyr$ by assuming that $L_{bol}$ is dominated
by the accretion luminosity
(i.e., $L_{acc} \sim G M_* \mdot/2R_* \sim 44 L_{\odot}$).
These parameters result in a maximum inner disk temperature
of $\sim 6600$ K, which should produce optical spectral features
analogous to that of an early-F stellar photosphere.
The reflected optical spectrum of V1647 Ori taken by Brice\~no et al. (2004)
shows Balmer absorption consistent with either an early-F or early-B
spectral type; those authors
favor the latter because of a lack of Ca II H \& K absorption.
However, the H \& K lines may be filled in by emission from
the wind or accretion flow, as has been seen in early spectra of
the FUor V1057 Cyg (Herbig 1977) and the prototype EXor EX Lupi
(Herbig et al. 2001).
A high-resolution spectrum would be required to resolve this issue.

After calculating the emergent flux from the disk, we applied extinction
assuming an interstellar extinction law and
$A_V=11$, similar to previous estimates based on the optical
spectrum of the source (Brice\~no et al. 2004) and the depth of
the 3$\mu$m ice feature (Vacca et al. 2004).
For the pre-outburst spectrum, we lowered the mass accretion rate
by a factor of $\sim$15, in accordance with the observed change in
bolometric luminosity.  This value is at the high end of the typical range
of values for T Tauri stars (Gullbring et al. 1998), and in particular
is similar to that of the flat-spectrum CTTS DG Tau (Gullbring et al. 2000)
and many Class I objects (Muzerolle et al. 1998; White \& Hillenbrand 2004).
We also increased the extinction to $A_V=15$,
consistent with the reddening change implied by the near-infrared
photometry (Reipurth \& Aspin 2004).
It is not clear whether this apparent extinction
change is the result of material near the central source suddenly
moving away or sublimating during the outburst, or simply the result
of a change in the scattered light component.  Since we do not include
the effects of scattered light in our model, we essentially ignore
this question by adopting an ad-hoc change in $A_V$.


The disk models provide excellent fits to the optical and near-infrared
portion of the pre- and post-outburst SEDs (Fig.~\ref{seds}),
especially given that the observations for each SED are not all simultaneous.
However, the mid- and far-infrared fluxes are much larger than
can be accounted for with the disk models alone.  Some of this discrepancy
could be explained by adopting a flared disk geometry, which would
increase the mid-infrared flux via increased reprocessing of light from
the central source.  Given the optical morphology of
the nebula and the large inferred extinctions, it is more likely
that there is an infalling envelope, heated by the central star
and accretion disk, that is producing most of the long-wavelength flux.

As a simple initial exploration, we have adopted an
optically thin remnant envelope model, similar to that calculated
by Adams et al. (1987) and Kenyon \& Hartmann (1991) to explain
FUor mid-infrared excesses.  There are several lines of evidence
that suggest the envelope around V1647 Ori may be tenuous.
One, the pre-outburst SED shows a flat spectrum, which is typically thought
to indicate a transitionary phase between Class I sources with dense
infalling envelopes and Class II sources with no envelope.
Two, millimeter and submillimeter observations put an upper limit
on the size of the emitting region of only $\sim 4000$ AU
(Lis et al. 1999; Mitchell et al. 2001; Andrews et al. 2004),
and Andrews et al. (2004) further estimate a circumstellar mass
of only $\sim 0.06 \; \msun$, which indicates that there is little left
of the original envelope.
In any case, the model is not intended
for deriving actual envelope parameters, but merely to illustrate
that the mid-infrared flux increase can be due
to the increased heating of envelope material by the larger $L_{acc}$.

The model assumes a spherically symmetric envelope with
finite inner and outer radii $R_{in}$ and $R_{out}$,
and a density distribution given by steady-state infall:
$\rho = {\mdot}_{inf}/[4{\pi}r^{3/2}(2G{M_*})^{1/2}]$,
where ${\mdot}_{inf}$ is the envelope mass infall rate
(e.g., Adams et al. 1987).  We adopted a temperature distribution
with a radial dependence $r^{-1/3}$, appropriate for optically thin
envelopes, as in Adams et al. (1987).  We used the proportionality
$T_{in} \propto L_{acc}^{1/6}$ given by Adams et al. (1987)
to scale the temperature at the inner radius between the pre-
and post-outburst models.  Finally, we assumed an envelope
filling factor $f < 1$ to approximate a bipolar cavity (the near
lobe of which is seen in the optical as McNeil's nebula).
The effects of inclination and scattering are ignored.

In order to produce the best fit to the pre-outburst SED,
we varied $R_{in}$, $R_{out}$, ${\mdot}_{inf}$, $T_{in}$, and $f$.
In general, the resulting values of these parameters
(see caption to Fig.~\ref{seds})
are physically plausible.
The temperature at the inner radius is similar to that of
the more detailed envelope models of Kenyon et al. (1993) at
similar radii.  The envelope outer radius is fairly typical
for Class I objects, and consistent with the observed upper limit
mentioned above.  The infall rate of $10^{-6} \; \msunyr$ is similar
to the expected value given typical gas temperatures
in molecular cloud cores such as in Taurus.  Finally, the filling factor
$f=0.3$ is consistent with a remnant envelope where most of
the material has already fallen onto the disk, while the accretion-powered
outflow has carved out a progressively larger cavity.
The total mass of the model envelope is $\sim 0.003 \; \msun$,
much less than the observed value, as expected if the observed mass
is dominated by the disk.
The same envelope model was used for the outburst spectrum, except
that the primary source of heating, the disk accretion luminosity,
was increased by the observed factor of $\sim 15$, and $T_{in}$ was
correspondingly increased by the proportionality given above.

Adding the envelope emission to the disk emission, we can qualitatively
explain both the pre- and post-outburst SEDs from the optical to 70 $\mu$m.
The outburst model slightly underpredicts the IRAC
measurements, which may indicate a hotter inner envelope temperature,
or possibly reprocessed light from a flared outer disk not taken into
account here.
Both models show a 10 $\mu$m emission feature, in contrast to most
flat-spectrum sources which usually show absorption.  An $N$-band
spectrum of V1647 Ori taken during the outburst by Andrews et al. (2004)
is featureless.  This may indicate that the actual envelope is not strictly
optically thin near 10 $\mu$m, or that the dust opacity is not typical
of the ISM, both likely possibilities.  A more detailed model taking these
effects into account is needed to understand the envelope
properties in detail.

\subsection{The Source of HH 22}

We have discovered a mid-infrared source coincident with HH 22A,
just $\sim 36"$ northeast of V1647 Ori.  This Herbig-Haro object is part of
HH 22, a series of knots seen in [SII] emission (Eisloffel \& Mundt 1997),
tracing an east-west jet perpendicular to the HH 23 outflow that is likely
emanating from V1647 Ori.  The knot HH 22A shows
faint extended emission at 1.3 mm, and considerably stronger HCO$^+$
emission than seen at the position of V1647 Ori (Lis et al. 1999).
The IRAC images show extended emission in a fan-shaped structure at
the position of HH 22A, aligned with the rest of the [SII] jet,
some other knots of which are seen in the IRAC 4.5 $\mu$m image.
We also see a possible ``counter-knot" to HH 22G, on the opposite side
and nearly equidistant from HH 22A, previously unseen.
The 24 {\micron} source is also coincident with HH 22A, and quite bright
($\sim 120$ mJy).  Our detections confirm that the HH 22 outflow
is produced by a separate protostellar object.
The source is extremely red, with an [8]-[24] color
of $\sim 7.2$.  Hence, it is likely an extremely embedded Class I source,
probably at an earlier evolutionary state than V1647 Ori itself.
                                                                                

\section{Discussion}

Our observations have shown that the outburst illuminating the optical
McNeil's nebula involved a brightness increase across the spectrum
to wavelengths as long as 70 $\mu$m.  The measured bolometric luminosity
increased by a factor of $\sim 15$ to 44 $L_{\odot}$.
Simple models of a viscous accretion disk can explain the optical
and near-infrared outburst emission, but an outer envelope component
heated by the accretion luminosity of the central source must be included
to explain the mid-infrared emission.
Assuming $L_{bol}$ is dominated by the accretion luminosity of the
central source, and assuming typical stellar parameters appropriate
for the pre-outburst luminosity, we estimate an approximate disk mass
accretion rate of $\sim 10^{-5} \; \msunyr$.  Such a rate is smaller
than typical of FUor variables by about a factor of 10 (Kenyon \&
Hartmann 1996).  However, we cannot reliably classify the outburst
with the present data.
Other properties of V1647 Ori, such as the optical and near-infrared
spectral features (Brice\~no et al. 2004; Vacca et al. 2004),
are something of an amalgam of
both FUor and EXor characteristics.
The true test will be in the ultimate timescale of the event.

Our results provide the first variability
measurements at such long wavelengths in young stellar objects.  
They show that outburst events may have a significant effect not only
on the evolution of the central star and disk, but also the envelope.
The additional heating and increased outflow activity may
contribute to the eventual clearing of envelope material, leading
to the onset of the Class II stage.
The timescale for the dust in the envelope
to respond to the change in luminosity of the central source should
be dominated by the light travel time, which for our model envelope size,
2000 AU, is roughly 5 days.  Since the outburst reached its peak
roughly a month before our observations, the outburst SED should
represent equilibrium conditions in the envelope.  Repeat observations
should be made as the outburst declines
in order to follow the evolution of the mid-infrared flux,
which will provide more detailed information on the envelope properties.

\acknowledgements

We thank N. Calvet, L. Hartmann, and C. Brice\~no for helpful discussions.
This work is based in part on observations made with the {\it{Spitzer Space
Telescope}}, which is operated by the Jet Propulsion Laboratory,
California Institute of Technology under NASA contract 1407.
Support for this work was provided by NASA through Contract Number
960785 issued by JPL/Caltech.

\clearpage

\begin{deluxetable}{lrrcccccc}
\tabletypesize{\small}
\tablecaption{{\it{Spitzer}} Fluxes ({\it{mJy}})}
\tablewidth{0pt}
\tablehead{
\colhead{Object} & \colhead{$\alpha$(J2000)} & \colhead{$\delta$(J2000)} &
\colhead{3.6$\mu$m} & \colhead{4.5$\mu$m} & \colhead{5.8$\mu$m} &
\colhead{8.0$\mu$m} & \colhead{24$\mu$m} & \colhead{70$\mu$m}}
\startdata
V1647 Ori & 05:46:13.1 & -00:06:04.8 & 2063.7 & 3089.0 & 3949.3 & 6427.3 & 15600.0 & 17600.0\\
HH 22 & 05:46:14.2 & -00:05:26.6 & 1.0 & 1.6 & 1.2 & 1.3 & 120.2 & \nodata \\
\enddata
\end{deluxetable}
                                                                                

\begin{figure}
\caption{IRAC 4.5 {\micron} (top) and MIPS 24 {\micron} (bottom)
images of V1647 Ori.  The source is extremely bright,
hence the visible PSF structure and saturation at the PSF core
of the 24 {\micron} source.
The T Tauri star LkH$\alpha$ 301 and
the newly-discovered source of the HH 22 outflow are also marked
for reference.
\label{24image}}
\end{figure}

\end{document}